# A radium assay technique using hydrous titanium oxide adsorbent for the Sudbury Neutrino Observatory

T.C. Andersen [a], R.A. Black [e], I. Blevis [b], J. Boger [c], E. Bonvin [d], M. Chen [b], B.T. Cleveland [e], X. Dai [e], F. Dalnoki-Veress [b], G. Doucas [e], J. Farine [b], H. Fergani [e], A.P. Ferraris [e], M.M. Fowler [f], R.L. Hahn [c], E.D. Hallman [g], C. K. Hargrove [b], H. Heron [e], E. Hooper [e], K.H. Howard [e], P. Jagam [a], N.A. Jelley [e1], A.B. Knox [e], H.W. Lee [d], I. Levine [b], W. Locke [e], S. Majerus [e], K. McFarlane [b], G. McGregor [e], G.G. Miller [f], M. Moorhead [e], A.J. Noble [b], M. Omori [e], J.K. Rowley [c], M. Shatkay [b], C. Shewchuk [b], J.J. Simpson [a], D. Sinclair [b], N.W. Tanner [e], R.K. Taplin [e], P.T. Trent [e], J.-X. Wang [a], J.B. Wilhelmy [f] and M. Yeh [c]

[a] *Physics Department, University of Guelph, Guelph, Ontario N1G 2W1, Canada*
[b] *Carleton University, Ottawa, Ontario K1S 5B6, Canada*
[c] *Chemistry Department, Brookhaven National Laboratory, Upton, New York 11973-5000, USA*
[d] *Department of Physics, Queen's University, Kingston, Ontario K7L 3N6, Canada*
[e] *Department of Physics, University of Oxford, Denys Wilkinson Building, Keble Road, Oxford, OX1 3RH, UK*
[f] *Los Alamos National Laboratory, Los Alamos, New Mexico 87545, USA*
[g] *Department of Physics and Astronomy, Laurentian University, Sudbury P3E 2C6, Canada*

**Abstract**

As photodisintegration of deuterons mimics the disintegration of deuterons by neutrinos, the accurate measurement of the radioactivity from thorium and uranium decay chains in the heavy water in the Sudbury Neutrino Observatory (SNO) is essential for the determination of the total solar neutrino flux. A radium assay technique of the required sensitivity is described that uses hydrous titanium oxide (HTiO) adsorbent on a filtration membrane together with a $\beta - \alpha$ delayed coincidence counting system. For a 200 tonne assay the detection limit for $^{232}$Th is a concentration of $\sim 3\times10^{-16}$g Th/g water and for $^{238}$U of $\sim 3\times10^{-16}$g U/g water. Results of assays of both the heavy and light water carried out during the first two years of data collection of SNO are presented.

## 1. Introduction

The Sudbury Neutrino Observatory (SNO) [1] is a 1000 tonne heavy water Cherenkov detector situated

[1] Corresponding author. Tel.: +11 441 865 273380; fax: +11 441 865 273418; e-mail: n.jelley1@physics.oxford.ac.uk

*Preprint submitted to Nuclear Instruments and Methods in Physics Research A*

at a depth of 6800 ft in INCO's Creighton mine at Sudbury, Canada. Its main aim is to observe neutrinos from the sun and investigate the origin of the observed deficit in their flux (the Solar Neutrino Problem). The detector uses ∼9500 photomultiplier tubes (PMTs) on a geodesic sphere of diameter 17.8 m to observe the Cherenkov light produced as a result of neutrino interactions occurring in the 1000 tonnes of $D_2O$. The $D_2O$ is held in a 12 m diameter acrylic sphere, which is surrounded by a shield of 7000 tonnes of $H_2O$ contained in a 34 m high barrel-shaped cavity of maximum diameter 22 m.

SNO can measure the flux of electron neutrinos from the sun via the charged current (CC) reaction: $\nu_e$ + d → p + p + $e^-$, and the total flux of all active neutrino flavours via the neutral current (NC) reaction $\nu_x$ + d → p + n + $\nu_x$, ($x$ = e, $\mu$, $\tau$). The first results from SNO on the total flux of all active neutrinos have recently been published [2], in which a comparison of the CC and NC flux provides evidence that neutrino flavour changing is the cause of the Solar Neutrino Problem.

## 2. Radioactivity limits in SNO

Of the naturally occurring radioactive nuclides, it is only in the decay chains of the two heavy nuclides, $^{232}$Th and $^{238}$U, that gammas with energies greater than 2.22 MeV can be produced. These can photodisintegrate the deuteron and hence mimic a NC event, as the signature of the NC reaction is the production of a free neutron. (At 6800ft depth, the number of cosmogenically-produced radioactive nuclides is negligible.)

The number of neutrons predicted by the Standard Solar Model that are produced by the neutral current interaction of solar neutrinos with deuterons is about 0.015 per day per tonne of heavy water. In the thorium decay chain $^{208}$Tl is produced with a branching ratio of 36% and in its decay a 2.614 MeV gamma is emitted with a 99.8% probability. The chance that this gamma photodisintegrates a deuteron is $2.07\times10^{-3}$, giving a probability that a neutron is generated of $7.45\times10^{-4}$ per Th decay. An upper limit on $^{232}$Th of $3.8\times10^{-15}$g/g (throughout this paper, g/g means gram of $^{232}$Th or $^{238}$U per gram of $D_2O$ or $H_2O$) corresponds to a photodisintegration background of one neutron per day per 1000 tonnes of heavy water. This level is less than 10% of the predicted NC rate and corresponds to an average Th decay rate of about one decay per day per tonne of heavy water.

In the uranium decay chain $^{214}$Bi is produced, and in its decay the probability that a gamma with an energy greater than 2.22 MeV is emitted is only ∼2%. The probability that a neutron is generated is $3.12\times10^{-5}$ per U decay and this sets a corresponding upper limit on $^{238}$U of $3.0\times10^{-14}$g/g. The light water surrounding the heavy water must also be very pure to reduce radioactivity backgrounds though the requirements are not so severe, with the concentration limits for $^{232}$Th and $^{238}$U, based on early simulations of the detector, being $37\times10^{-15}$g/g and $45\times10^{-14}$g/g, respectively.

To achieve this level of radiopurity, both the heavy and light water in SNO were purified via several stages including filtration, ion exchange and reverse osmosis, which are described in detail in [1]. The water can be recirculated for further purification and assay. All the materials used in the construction were also selected for low radioactivity. To reduce any radon gas contamination of both the light and heavy water, all components in the water systems were chosen for low radon diffusion and emanation and the surfaces of the $D_2O$ and $H_2O$ are covered with $N_2$ gas obtained from liquid nitrogen.

It is essential to be able to assay the heavy water in the SNO detector at these very low levels of radioactivity to check that the required radiopurity has been achieved and measure the NC background. With this aim in mind, the hydrous titanium oxide (HTiO) ion-exchange system was developed to assay Ra, Th and Pb isotopes. It is one of four complimentary techniques used in SNO: the other three being an $MnO_x$ assay method, for Ra isotopes, degassing for $^{222}$Rn from the U chain, and direct counting of the amount of Cherenkov light from the decay of $^{208}$Tl and $^{214}$Bi. These other techniques will be described elsewhere.

## 3. The HTiO assay method

In this technique heavy or light water is passed through HTiO trapped on the filtration fibres of microfiltration membranes with a pore size of 0.1$\mu$m.

(Hydrous titanium oxide is a non-stoichiometric compound so the chemical formula HTiO is only representational.) Hydrous titanium oxide is an inorganic ion-exchanger [3,4] that has been used in the nuclear industry for the extraction of heavy ions. It is formed as a white colloidal suspension by the hydrolysis of titanium salts [5], which forms particles of sizes ranging from 1 to 100 microns, with a mode at around 10 microns. Examples of its use are given in [6,7,8]. The initial investigations for SNO on the use of HTiO for extracting radium, thorium and lead are described in [9,10,11]. In this paper the extraction and assay of radium from the Th and U chains is reported.

In the HTiO assay procedure there are five steps: deposition, extraction, elution, secondary concentration and counting. The HTiO ion-exchanger is first deposited onto microfiltration membranes: 0.25 m long filters for a light water assay and 1.0 m long filters for a heavy water assay. Then the columns containing the loaded filters are taken underground to extract $^{224}$Ra from the Th chain and $^{226}$Ra from the U chain from a large volume of $D_2O$ (∼200 tonnes) or $H_2O$ (∼30 tonnes). After extraction, the columns are brought back up to the surface laboratory, and the radium is eluted with 15 *l* of dilute nitric acid. In the secondary concentration stage, the radium in the 15 *l* nitric eluate is further concentrated down to ∼10 m*l* of final eluate. This is then mixed with ∼40 g of liquid scintillator and counted using a $\beta-\alpha$ delayed-coincidence liquid-scintillation counter.

The measurement of the concentration of $^{224}$Ra and $^{226}$Ra therefore requires the determination of the extraction efficiency of the loaded filters for a throughput of ∼200 tonnes of water, the subsequent elution and secondary concentration efficiencies, and finally the detection efficiency of the $\beta-\alpha$ counters. The inference of the concentration of the elements in the Th and U chain in the heavy or light water then requires knowledge of the equilibrium conditions within the chains, and of the extent of any radioactive species plating out within the water systems.

Initially only heavy water assays were made, using hollow fibre 1.0 m filters, and then the assays were extended to both the light and heavy water using 0.25 m and 1.0 m pleated membrane filters, respectively. In this paper the determination of the extraction and elution efficiencies for both these filters are first explained; then the $\beta-\alpha$ counting system and its detection efficiency are described; finally the assay results are presented and discussed.

### 3.1. *Production of HTiO*

The HTiO is produced by adding 400 m*l* of 15% w/v $Ti(SO_4)_2$ to 3 *l* of ultrapure water (UPW) and then adjusting to pH 12 with 3.5 *l* of 0.5 M NaOH. After allowing the solution to sit in a cool dark area for one week, it is rinsed with UPW to reduce the $Na^+$ and $SO_4{}^{2-}$ contaminants. The first step in rinsing the HTiO involves reducing the volume of the undisturbed HTiO solution to ∼5 *l* by aspirating the supernatant. The remaining solution is then stirred for 5 minutes and equally divided into eight centrifuge bottles and centrifuged at 3400 rpm for 7 minutes. After aspirating and discarding the supernatant, the precipitate is combined with ∼600 m*l* of UPW and stirred for 5 minutes. The centrifuging and rinsing steps are repeated two more times. In the final step, the precipitates in the eight centrifuge bottles are combined in an amber bottle and the weight of the HTiO solution is made up to 5 kg with UPW.

### 3.2. *Hollow fibre filter system*

Before 2001, only heavy water assays were made using two hollow fibre 1.0 m Amicon H26 filters in parallel, each with a pore size of 0.1$\mu$m and a surface area of 2.4 m$^2$. About 1.25 g Ti in the form of HTiO was deposited onto each filter, and the coverage was therefore 0.5 g Ti/m$^2$. A specially constructed deposition and elution rig in the surface laboratory was used for this purpose. Because of the need to keep radioactivity backgrounds to a minimum this rig is made principally out of low activity polypropylene.

The HTiO is deposited by adding the required amount of concentrated HTiO stock to a reservoir on the elution rig. The volume is then made up to 15 *l* with UPW and circulated through the filter at ∼85 *l*/min for 10 minutes.

To determine the radium background from the HTiO, the chemicals used in the elution and secondary concentration, the filters and the elution rig, the loaded filters are first eluted, without any extraction of radium from the heavy water, and the eluate concentrated and counted. The same filters are then cleaned by dissolving the HTiO in 0.5 M HCl, by circulating the acid for forty minutes and leaving the filters to soak overnight. This is followed by a UPW rinse before reloading the filters with HTiO from the same batch. The filters are then taken underground.

Table 1
Summary of efficiencies for the assay of radium

| Efficiency | Hollow | Pleated |
|---|---|---|
| Extraction $\varepsilon_{ext}$ | 95±5% | $95^{+5}_{-10}$%[a] |
| Elution $\varepsilon_{elu}$ | 70±5% | 80±10% |
| Secondary conc $\varepsilon_{conc}$ | 43±7% | 71±10% |
| H1 extraction | 95±5% | - |
| H1 elution | 53±5% | - |
| MediaKap-10 extraction | 95±5% | 95±5% |
| MediaKap-10 elution | 90±10% | 75±10% |
| Chemical $\varepsilon_{ext}\cdot\varepsilon_{elu}\cdot\varepsilon_{conc}$ | 29±5% | 48±11% |
| Counting $\varepsilon_{count}$ | | |
| Th chain | 45±5% | 45±5% |
| U chain | 60±10% | 60±10% |
| Total $\varepsilon_{ext}\cdot\varepsilon_{elu}\cdot\varepsilon_{conc}\cdot\ \varepsilon_{count}$ | | |
| Th chain | 13±2% | 24±5% |
| U chain | 17±4% | 32±8% |

[a] This is for heavy water; extraction efficiency for light water assays is 96±4%, see text.

Once underground the two filters must be deuterated to maintain the isotopic purity of the heavy water. The two filters are first slowly filled from the bottom with ∼10 kg of $D_2O$ over 2 hours, displacing any $H_2O$ out of the top. The filters are left overnight to allow time for $H_2O$ deep within the filter to diffuse out into the surrounding $D_2O$ and the flow is resumed the next morning. After ∼9 hours flow and ∼20 kg more $D_2O$, the density of the water coming out of the top of the filters is 1.102-1.103, and the deuteration of the filters is then stopped (the density of pure $D_2O$ is 1.105 g/cc at 20 $^0$C).

#### 3.2.1. *Extraction efficiency*

The extraction efficiency for a loaded filter depends on the volume of water passed per unit area, the purity of the water, the flow rate per unit area, the coverage of the adsorbent (g/m$^2$) and on the distribution or partition coefficient of the adsorbent ($K_d$). Provided the volume of water that has passed through the filter is such that the ratio of the amount of activity on the adsorbent to its concentration in the liquid is much less than $K_d$, then the extraction efficiency will be close to 100%. Other contaminants in the water, if of sufficient concentration, can reduce the efficiency of the adsorbent if large volumes of water are processed.

Measurements of the extraction efficiency of the hollow fibre filters were carried out at Oxford [10] using radium spikes. A small Amicon H1 filter was used of area 0.015 m$^2$ and with the same pore size and with the same coverage of HTiO as for the large 1.0 m filters. Using the same flow rate per unit area, these tests indicated that the extraction efficiency was ∼95%, even for an equivalent processed volume of 1 kilotonne (≡ 200 tonnes/m$^2$)and flow rate per unit area of ∼40 *l*/min/m$^2$. Typical values for the heavy water assays with the hollow fibre filters were 200 tonnes (≡ 40 tonnes/m$^2$) at 20 *l*/min/m$^2$. At the site of the experiment, stable barium, for which HTiO has a similar but slightly reduced affinity [13] compared to radium, was added to check the extraction efficiency after an assay and the barium extraction efficiency was found to be ∼90%.

These measurements give an extraction efficiency with the 1.0 m hollow fibre filters of 95±5% for Ra using the Ba data as the lower bound and 100% as the upper bound.

#### 3.2.2. *Elution and secondary concentration efficiencies*

The elution efficiencies for the heavy water assays using the hollow fibre filters are based on $^{226}$Ra spike experiments that were carried out on 0.25 m Memtrex pleated membrane filters at Laurentian University. It is assumed that the elution efficiency depends on the acid strength and contact time and further, because the efficiency depends on the binding ability of Ra with HTiO, that there is no significant difference in the elution efficiencies for the hollow fibre and pleated membrane filters with the same contact time. A 20 minute circulation of 15 *l* of 0.03 M $HNO_3$ was used for the hollow fibre filters, which achieved an elution efficiency of 70%. For comparison, the elution efficiency of Ba, which would be slightly higher than that of Ra due to the relatively weaker affinity of Ba to HTiO, was found to be 75±10%.

The secondary concentration of the 15 *l* nitric eluate was achieved by first adding ∼85 mg of Ti in the form of $Ti(SO_4)_2$ solution, adjusting the pH to 9 by titration with NaOH to form HTiO, with which the Ra co-precipitates. The 15 *l* solution was then filtered through a small Amicon H1 filter to trap the precipitated HTiO+Ra, and the Ra eluted off the H1 filter by using 140 m*l* of 0.03M $HNO_3$ twice for 20 minutes each. About 50 mg of Ti in the form of $Ti(SO_4)_2$ was added to the combined $HNO_3$ eluate. The pH was adjusted to 9 by titration with NaOH to form HTiO with which the Ra co-precipitates. Then the titrated eluate was passed through a MediaKap-10 filter to extract the Ra. Finally the Ra was concentrated down to ∼10 m*l* of acid solution by eluting the radium off the MediaKap-10 filter with 5 m*l* of 0.5 M HCl used twice.

The efficiency of this secondary concentration was determined by $^{224}$Ra and $^{226}$Ra spike experiments, carried out on H1 and MediaKap-10 filters under the same conditions as in the actual assay. The final figure of 43±7% is the product of the extraction and elution efficiencies of the H1 filter and MediaKap-10 filters that are used in this stage of the chemical procedure (see Table 1 for details).

### 3.3. *Pleated membrane filter*

From 2001, two 1.0 m Memtrex pleated membrane filters [14] (Osmonics, Inc., USA) in parallel were used instead of the hollow fibre filters as the pleated membrane filters had a higher pressure rating, which would allow higher flow rates to be used. The pleated membrane filter has 0.52 $m^2$ of surface area for a 0.25 m filter, and therefore 2.08 $m^2$ for a 1.0 m filter, with a 0.1 $\mu$m pore size. The filters are cylindrical (diameter 2.7″) with a double O-ring seal at one end, and are held in a cylindrical holder (the column). The water enters at one end of the column, flows around the outside and through the HTiO loaded membrane into the inside of the filter, and out the other end of the column.

Several modifications, mainly in the secondary concentration stage where the H1 filter is replaced with three MediaKap-10 filters in parallel, were made to improve the chemical procedural efficiency.

#### 3.3.1. *Extraction efficiency*

To examine the full-scale extraction efficiency, two kinds of experiments were conducted: one using the underground water systems and the other, the facility at Carleton University where the extraction of radioactive spikes at the same flow rate per unit area used underground but with a lower mass/$m^2$ of $\sim$ 1 tonne water/$m^2$ could be measured.

Two light water assays, LWA2 and LWA3, were first carried out to examine the extraction in the actual underground light water system. In these experiments two 0.25 m filters with HTiO coverage of 0.5 g Ti/$m^2$ were placed underground in series to process about 47 tonnes (≡90 T/$m^2$) of light water at the flow rate of about 18 *l*/min (≡35 *l*/min/$m^2$). The radioactivity extracted by these two columns was then separately measured. If the extraction efficiency is high, there should be little radioactivity on the downstream column. However, in both experiments, significant activity was found to penetrate the upstream filter and reach the downstream one (see Table 2). It was suspected that the HTiO coverage on the pleated membrane filter might be too low, either to provide complete coverage or to cope with contaminants in the light water system, or that the high water flow rate might result in incomplete extraction.

To solve this problem, a heavier HTiO loading (1.25 g Ti in the form of HTiO, equivalent to 2.5 g Ti/$m^2$) and a slower flow rate ∼10 *l*/min (≡19 *l*/min/$m^2$) were used in the light water assays LWA4-8. As a result, the average radium extraction of the upstream filter was improved to $80 \pm 4$%, and for the largest tonnes water/$m^2$ processed, LWA8, the efficiency was $80 \pm 5$%. To improve the signal in the light water assays, the activity on both the upstream and downstream filters was added together, which gives an overall extraction efficiency of $96 \pm 4$%, taking the efficiency of both filters as 80%.

After an extraction a loaded filter was examined to check the HTiO coverage. The passage of a large volume of water had not caused the coverage of HTiO to become non-uniform and the extraction efficiency of 80% for the upstream pleated filter, rather than 95% as found for the hollow fibre filter, is probably caused by some suspended non-electrolytes in the light water. These would cover the exchange sites of the HTiO adsorbent and therefore reduce the exchange capacity of the HTiO. The level of such suspended particles in the heavy water is expected to be much lower, because the heavy water is in contact with far less and cleaner material than the light water. (The level of any inorganic contaminant in the light water is less than 1 ppb and the level of organics in $H_2O$ is also low and less

Table 2
The extraction efficiency for pleated membrane filters in the underground water systems

| Expt | T/m$^2$ | *l*/min /m$^2$ | g Ti /m$^2$ | Upstream filter[a] ($10^{-15}$g/g) | | Downstream filter[a] ($10^{-15}$g/g) | | Extraction Efficiency[b] (%) | |
|---|---|---|---|---|---|---|---|---|---|
| | | | | Th | U | Th | U | $^{224}$Ra | $^{226}$Ra |
| LWA2 | 88 | 35 | 0.5 | $19.2 \pm 2.7$ | $0.94 \pm 0.21$ | $17.5 \pm 2.4$ | $0.99 \pm 0.26$ | - | - |
| LWA3 | 90 | 35 | 0.5 | $22.0 \pm 2.6$ | $1.87 \pm 0.76$ | $11.8 \pm 2.5$ | $0.79 \pm 0.74$ | - | - |
| LWA4 | 21 | 19 | 2.5 | $22.1 \pm 1.1$ | $8.5 \pm 5.2$ | $0^{+6.2}_{-0}$ | $2.0^{+4.6}_{-2.0}$ | $100^{+0}_{-28}$ | $76^{+24}_{-56}$ |
| LWA5 | 36 | 19 | 2.5 | $60.1 \pm 5.4$ | $3.73 \pm 0.90$ | $18.6 \pm 4.2$ | $1.11 \pm 0.79$ | $69 \pm 8$ | $70 \pm 22$ |
| LWA6 | 66 | 19 | 2.5 | $47.7 \pm 3.8$ | $4.37 \pm 1.25$ | $10.4 \pm 3.0$ | $0.61^{+0.86}_{-0.61}$ | $78 \pm 6$ | $86^{+14}_{-20}$ |
| LWA7 | 53 | 19 | 2.5 | $71.3 \pm 3.9$ | $2.79 \pm 0.42$ | $13.5 \pm 3.0$ | $0.33^{+0.34}_{-0.33}$ | $81 \pm 4$ | $88 \pm 12$ |
| LWA8 | 82 | 19 | 2.5 | $47.4 \pm 2.9$ | $2.22 \pm 0.47$ | $9.7 \pm 2.5$ | $0.43^{+0.48}_{-0.43}$ | $80 \pm 5$ | $81 \pm 19$ |
| RDSA11 | 64 | 19 | 2.5 | $1.07 \pm 0.14$ | $0.28 \pm 0.09$ | $0^{+0.17}_{-0}$ | $0.01^{+0.09}_{-0.01}$ | $100^{+0}_{-16}$ | $96^{+4}_{-31}$ |

[a] The equivalent Th(U) concentration at the end of the extraction.
[b] Extraction efficiency = 1 – Downstream Th(U)/Upstream Th(U).

than 30 ppb. These levels are unlikely to reduce the extraction ability of the HTiO for any activity.)

To measure the extraction efficiency for heavy water, an assay (RDSA11) of the salinated heavy water (0.2% NaCl) was made using two pairs of 1.0 m pleated membrane filters in series. To determine the radium background, each pair was first loaded with HTiO which was then eluted, the eluate concentrated and the background radium activity counted. The filters were then cleaned by circulating 0.5 M HCl, reloaded with HTiO and taken underground for the assay. The results for this assay are given in Table 2 and show that the extraction efficiency of the pleated membrane filter is $100^{+0}_{-16}$% for 64 tonnes of water/m$^2$ and a flow rate of 19 *l*/min/m$^2$.

In this experiment the $^{224}$Ra background of the upstream pair was anomalously high, indicating some contamination (probably from the elution rig). The cleaning of the filter would have cleaned the elution rig as well and so the average rate from four background runs has been assumed for the upstream value of the $^{224}$Ra background. The extraction efficiency is in agreement with $^{224}$Ra spike experiments that were conducted at Carleton University to check the extraction efficiency of radium in 0.5% NaCl solution using a single 0.25 m filter with a loading of 2.5g Ti/m$^2$. Satisfactory extraction efficiencies (94%, 97%, 94%) were obtained at a flow rate/m$^2$ of 23 *l*/min/m$^2$ with water amounts of 0.35, 0.58 and 0.92 tonne/m$^2$, respectively.

Combining these results, the extraction efficiency of the pleated membrane filter is taken to be $95^{+5}_{-10}$% for all assays of SNO heavy water, whether salinated or not.

### 3.3.2. *Elution and secondary concentration efficiencies*

After an assay the columns are transported to the surface laboratory and mounted on the elution rig. To improve the elution efficiency a 40 minute circulation of 15 *l* of 0.03 M $HNO_3$ is used, which achieves an elution efficiency of 80%, compared to the 70% achieved previously for the hollow fibre filters with a 20 minute circulation.

The secondary concentration is carried out by first adding $Ti(SO_4)_2$ and then NaOH, as with the hollow fibre filters, to form HTiO with which the Ra coprecipitates. Then, in order to improve the secondary concentration efficiency, the H1 filter followed by a MediaKap-10 filter, used with the hollow fibre filters, is replaced with three MediaKap-10 filters in parallel through which the 15 *l* solution is then filtered. The radium is eluted from each with 3.5 m*l* 0.2 M $HNO_3$ for 15 minutes. These are combined to form the first ~10.5 m*l* eluate. The MediaKap-10 filters are eluted twice more to ensure all the activity is collected, and combined to form the second and third ~10.5 m*l* eluates.

The elution and secondary concentration efficiencies were obtained through $^{226}$Ra and $^{228}$Th spike experiments at Laurentian University. For the elution the same acid strength and contact time were used as in an assay, and for the secondary concentration the conditions were the same as in actual assays.

The elution and secondary concentration efficiencies are estimated to be, respectively, 80±10% and 71±10% for radium. (MediaKap-10 extraction efficiency is 95±5% and the elution efficiency is 75±10%, see Table 1). The elution efficiencies for $^{228}$Th and $^{212}$Pb were determined by $^{228}$Th spike experiments to be very low (1.6% and 5%, respectively).

$^{232}$Th $\alpha$ $1.4\times10^{10}y$
$^{228}$Ra $\beta$ $5.8y$
$^{228}$Ac $\beta$ $6.2h$
$^{228}$Th $\alpha$ $1.9y$
$^{224}$Ra $\alpha$ $3.7d$
$^{220}$Rn $\alpha$ $56s$
$^{216}$Po $\alpha$ $0.15s$
$^{212}$Pb $\beta$ $10.6h$
$^{212}$Bi $\alpha$ 36% $\beta$ 64% $61m$
$^{208}$Tl $\beta$ $3m$ $^{212}$Po $\alpha$ $0.3\mu s$
$^{208}$Pb

$^{238}$U $\alpha$ $4.5\times10^{9}y$
$^{234}$Th $\beta$ $24d$
$^{234}$Pa $\beta$ $1.2m$
$^{234}$U $\alpha$ $2.5\times10^{5}y$
$^{230}$Th $\alpha$ $7.5\times10^{4}y$
$^{226}$Ra $\alpha$ $1600y$
$^{222}$Rn $\alpha$ $3.8\,d$
$^{218}$Po $\alpha$ $3\,m$
$^{214}$Pb $\beta$ $27m$
$^{214}$Bi $\alpha$ 0.02% $\beta$ >99% $20m$
$^{210}$Tl $\beta$ $1.3m$ $^{214}$Po $\alpha$ $164\mu s$
$^{210}$Pb $\beta$ $22y$
$^{210}$Bi $\beta$ $5d$
$^{210}$Po $\alpha$ $138d$
$^{206}$Pb

Fig. 1. Thorium and Uranium decay schemes

### 3.4. $\beta-\alpha$ *counter and counting efficiency*

For a 200 tonne assay of heavy water with concentrations of $^{232}$Th of $3.8\times10^{-15}$g/g and of $^{238}$U of $3.0\times10^{-14}$g/g, corresponding to background neutron production rates from deuteron photodistintegration of ∼10% of that predicted by the Standard Solar Model, the radium decay rates, if 50% is extracted, eluted and concentrated, are ∼130 decays per hour (dph) for $^{226}$Ra , while only ∼6 dph for $^{224}$Ra. This very low level of activity requires counters with high efficiency and low background. To meet these requirements, a $\beta-\alpha$ counting system was developed in Oxford [10], which uses delayed coincidence between two sequential decays, which follow the decay of Ra and occur towards the end of the Th and U chains, to reduce the background, together with pulse shape discrimination [16]. The decay schemes are shown in Fig. 1.

For the Th chain, the $\beta$-decay of $^{212}$Bi and the $\alpha$-decay of the $^{212}$Po are utilized, while for the U chain it is the $\beta$-decay of $^{214}$Bi and the $\alpha$-decay of $^{214}$Po. The different half-lives of the two Po isotopes (300 ns for $^{212}$Po and 164 $\mu$s for $^{214}$Po) allow these $\beta-\alpha$ sequential decays to be easily identified and separated from each other. In previous $\beta-\alpha$ coincidence techniques [17,18], the activity was deposited as a thin film. However, such techniques have quite low efficiencies and producing thin films from HTiO eluates was found to be difficult. To overcome these limitations, the activity is concentrated down to a small volume of aqueous solution, which is then mixed into a liquid scintillator and the $\beta-\alpha$ decays counted with a photomultiplier.

Plastic jars with their high radiopurity and low cost proved the ideal disposable container for the $\beta-\alpha$ counting. After testing various combinations [10], the final procedure chosen is that a ∼10.5 m*l* aqueous sample containing the eluted radium is mixed with 42 g of Optiphase Hi-Safe 3 liquid scintillator and placed in a 60 m*l* polymethylpentene jar. The 5 cm diameter jar and a 5 cm photomultiplier (an Electron Tubes Ltd 9266XB PMT) are optically coupled with silicone grease to maximise light collection. The counters are shielded against soft room gammas by a 2.5 cm thick oxygen-free, high conductivity copper housing. The system currently uses eight of these counters.

The electronics block diagram is shown in Fig. 2. All events with coincidences within a time window $<700\ \mu$s are captured by a CAMAC system and their TDC and 3 ADC values (beta full-pulse, alpha full-pulse and alpha tail-pulse) are transferred to a PC. The ADCs are operated as QDCs and the delay for the alpha tail is set so that the integrated charge in the tail of the pulse is approximately a third of that in the full alpha pulse.

A time cut on the time T between the $\beta$ and the $\alpha$ signal is first used to tag an event as thorium-like (T $<1500$ ns, corresponding to the 300 ns half-life of $^{212}$Po) or uranium-like (10 $\mu$s$<$ T $<700$ $\mu$s, corresponding to the 164 $\mu$s half-life of $^{214}$Po). In order to improve the background rejection from random coincidences, three off-line software cuts are applied, illustrated in Fig. 3. A $\beta$ energy cut, which is normally fixed at 75 keV $\lesssim\beta\lesssim$ 2500 keV, removes cross-over coincidences between different counters. Alpha energy (6 MeV $\lesssim \alpha \lesssim$ 14 MeV) and PSD (pulse shape discrimination, which is the ratio of $\alpha$ tail charge/ full charge) cuts greatly help to reduce the background events from random $\gamma-\gamma$ coincidences. These cuts are set using spiked calibration sources. Different measurement conditions, such as temperature and cloudiness of sample, can lead to a slight $\alpha$ peak shift. Therefore, $\alpha$ and PSD cuts are adjusted by software for every individual sample. In Fig. 3, it can be seen that the true and background events can be easily distinguished.

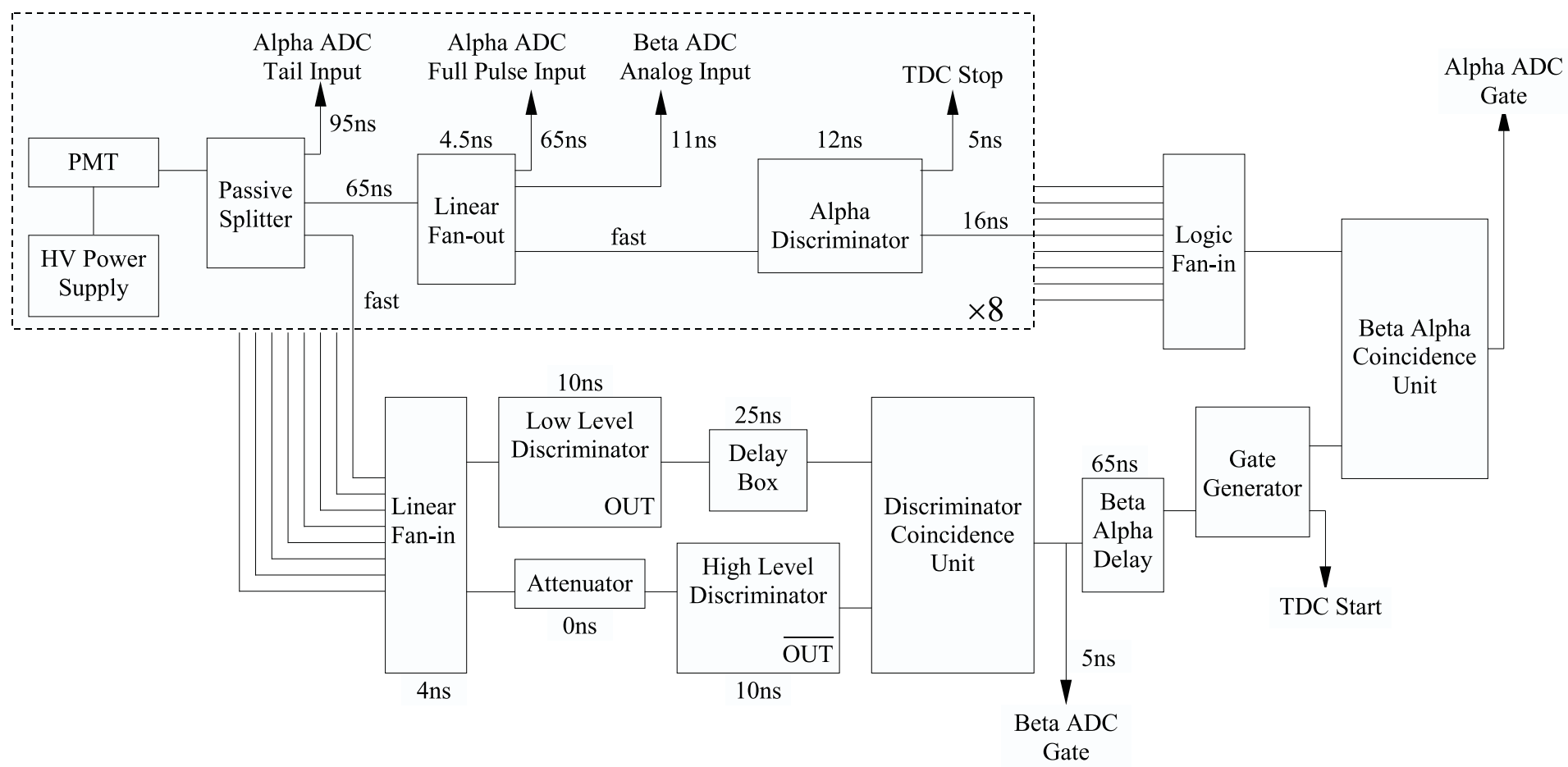

Fig. 2. Electronics block diagram of the $\beta-\alpha$ counters.

A good linear relationship between counting rate and sample activity was found using blank and spiked sources of order 0.1 mBq to 1 Bq of $^{228}$Th and $^{226}$Ra, and the blank source (a mixture of 10 m*l* 0.5 M HCl and 42 g scintillator) gave $\sim$0.03 cph (counts per hour) for the Th chain and $\sim$0.3 cph for the U chain activities. These count rates are much lower than the the total radium backgrounds using the pleated membrane filters for both the Th and U chains of typically 0.50 and 1.34 cph, respectively, which correspond to $\sim 8\times10^{-16}$g Th/g and to $\sim 5\times10^{-16}$g U/g for a 200 tonne assay.

Efficiency calibrations were also made. The result of many calibrations of similar sets of counters gave counting efficiencies of 45$\pm$5% for the Th chain and 60$\pm$10% for the U chain. In addition, as $^{228}$Th ($\sim$1 Bq) and $^{226}$Ra ($\sim$0.5 Bq) sources are regularly used to check the performances of the counters, any change in the counting efficiencies can easily be seen. In some of our heavy water assays, a few samples had a light yellow colour that might be attributed to manganese and organic matter in the water. To check for any loss in efficiency, small $^{212}$Pb spikes were added to the yellow samples after counting; no significant loss in counting efficiency was found.

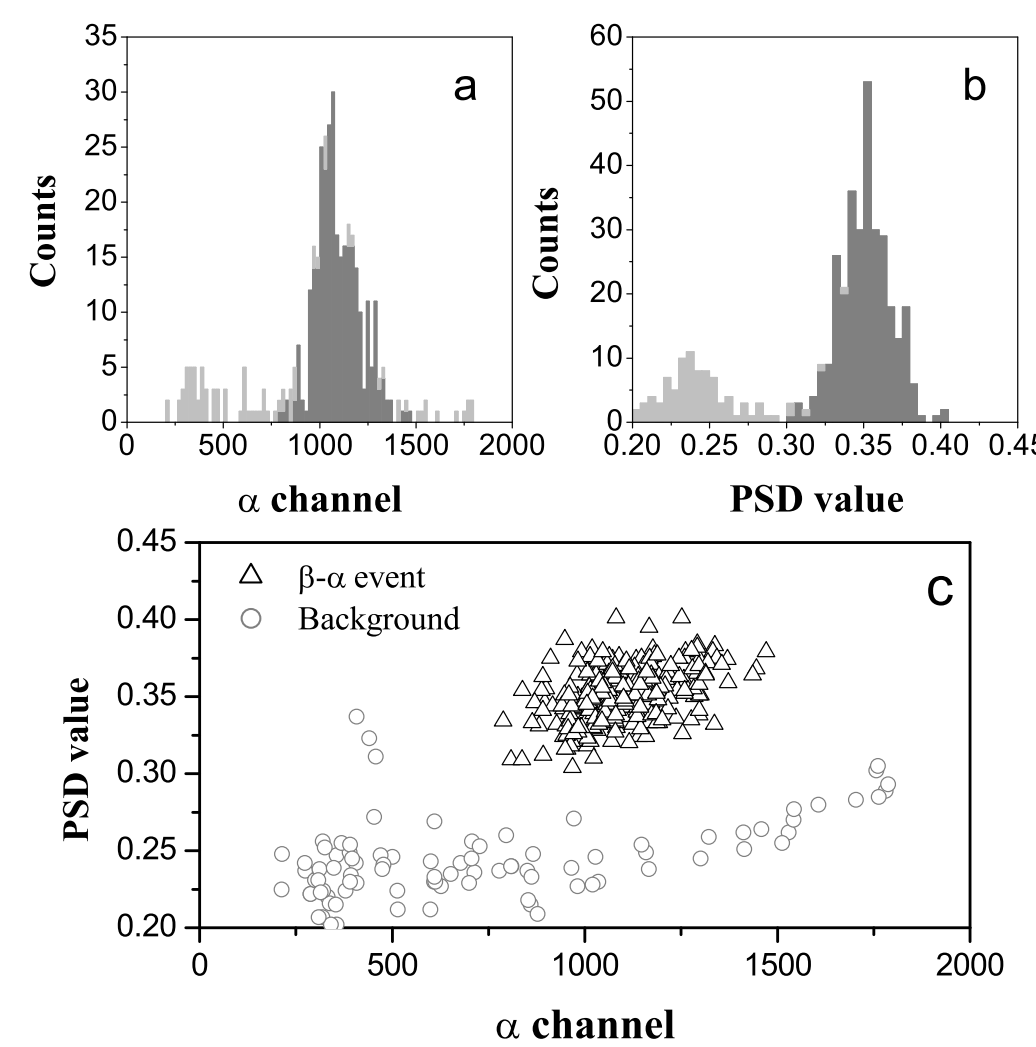

Fig. 3. Thorium chain $\beta-\alpha$ coincidence data for assay RDA5. a: The full charge spectrum for the alpha (second) pulse of the coincidence. b: The spectrum for the PSD ratio of the tail to the full charge of the alpha. c: Two dimensional PSD$-vs-$alpha spectrum. The $\beta-\alpha$ events pass cuts on the full alpha charge and the PSD ratio and are plotted darker than the background events in all three spectra.

## 4. Description of assay experiments

The experiments included in this paper are named according to a code system which is: RDA# (Radium $D_2O$ Assay), RDSA# (Radium $D_2O$ Salt Assay- an assay of the $D_2O$ with 0.2% NaCl added), LWA# (Radium $H_2O$ assay). From October 1999 to November 2001, nine heavy water and eight light water assays have been conducted. Two 1.0 m filters in parallel were used for all heavy water assays, two 0.25 m filters in series for all light water assays. The hollow fibre filters were used for RDA2, RDA3, RDA4 and RDA5. For the other assays, the pleated membrane filters were used.

The accurate measurement of the amount of radium in the water volume also requires the determination of any contribution from (a) the pipes connecting the acrylic vessel and cavity to the filters, and from (b) the filters, the elution rig, the secondary concentration apparatus, the chemicals and the counters. To determine (a) an assay was carried out of water which was circulated through the flow and return pipes to the acrylic vessel but with the vessel by-passed. The result was a limit for Th of less than $0.6\times10^{-15}$g/g [12]. To determine (b) a background measurement using the same filter is always carried out a few days before an assay with the same procedure but no extraction step.

To minimise the procedural background, the filters, the counting jars and the tubing used in the secondary concentration are only used once for each assay. The titanium concentration in all solutions produced in the entire procedure was measured by inductively coupled plasma mass spectrometry and UV-visible spectrometry to check for any anomalous Ti loss; none was found. Since May 2001 (assay LWA4 and subsequent assays), the HTiO coverage on the pleated membrane filters has been increased from 0.5 to 2.2-2.5 g Ti/m$^2$.

The heavy water assays sampled the $D_2O$ by drawing water from the bottom of the acrylic vessel and returning it to the top. The light water assays sampled the $H_2O$ between the acrylic vessel and the photomultipliers, where the radioactivity contributes most to the background of the neutrino-induced signal.

### 4.1. *Data Reduction*

In the Th chain (Fig. 1), the rate of delayed coincidences is directly proportional to the specific activity of $^{212}$Bi, $\lambda_{\rm Bi}N_{\rm Bi}$, where $\lambda$ is the decay constant, and $N$ is the number of atoms. If the radionuclides above $^{212}$Bi are not in secular equilibrium, then the rate of coincidences varies with time as the chain equilibrates. A spectrum of the counting rate plotted against time can be fitted to competing exponentials, and the specific activity of isotopes further up the chain can be derived. In this fashion the activity levels of $^{228}$Th, $^{224}$Ra, $^{212}$Pb and $^{212}$Bi can be measured. Likewise, in the U chain (Fig. 1), the rate of $^{214}$Bi decays, plotted against time, can be used to fit the activity levels of $^{226}$Ra, $^{222}$Rn, $^{214}$Pb and $^{214}$Bi. The fitting program uses a log likelihood method [10] with Poisson distributed bin heights.

An example of spectrum fitting is shown in Fig. 4. The total $\beta-\alpha$ coincidence rate is given with the contributions from different radionuclides in the chain. For the Th chain data in Fig. 4, the curve labelled $^{224}$Ra shows the contribution from the activity of $^{224}$Ra alone; initially the $\beta-\alpha$ coincidence rate builds up in a time determined by the $^{212}$Pb lifetime (10.6 hr) to a maximum 37 hr after the start of counting. It then decays away at a rate determined by the half life of $^{224}$Ra (3.66 d). For the U chain data shown in Fig. 4, it should be noted that $^{214}$Pb and $^{214}$Bi can not be separated very well due to their close half-lives (26.8 min for $^{214}$Pb and 19.9 min for $^{214}$Bi), and their contribution is labelled $^{214}$Pb/$^{214}$Bi. Thorium, radium and lead isotopes can all be extracted with high efficiencies from the underground water system by the HTiO ion-exchanger [10]. However, at the elution stage, very little thorium and lead are eluted by the 0.03M $HNO_3$ solution. Thus, almost no $^{228}$Th (which is above $^{224}$Ra in the decay chain) can be found in the final counting samples, as illustrated in Fig. 4.

During the time interval between the end of the elution and the beginning of counting, which was normally 25 hours for the hollow fibre procedure and 5 hours for the pleated membrane procedure, the Pb and Bi isotopes will build up from the Ra isotopes. That is why contributions from the Pb and Bi isotopes are seen at the beginning of the counting period. (In the U chain their activities are relatively high in both background and assay spectra, because of $^{222}$Rn contamination from exposure of the acidic eluate and liquid scintillator to the air.) Because of their short half-lives, Pb and Bi isotopes will decay away very fast and do not significantly affect the fitting accuracy for the Ra isotopes. The calculated activities (extracted plus background) of the relevant isotopes at the beginning of the counting are listed above Fig. 4. For the RDA5 assay with hollow fibre filters, the background counting rates were 0.15 cph and 0.55 cph for the Th and U chain, respectively.

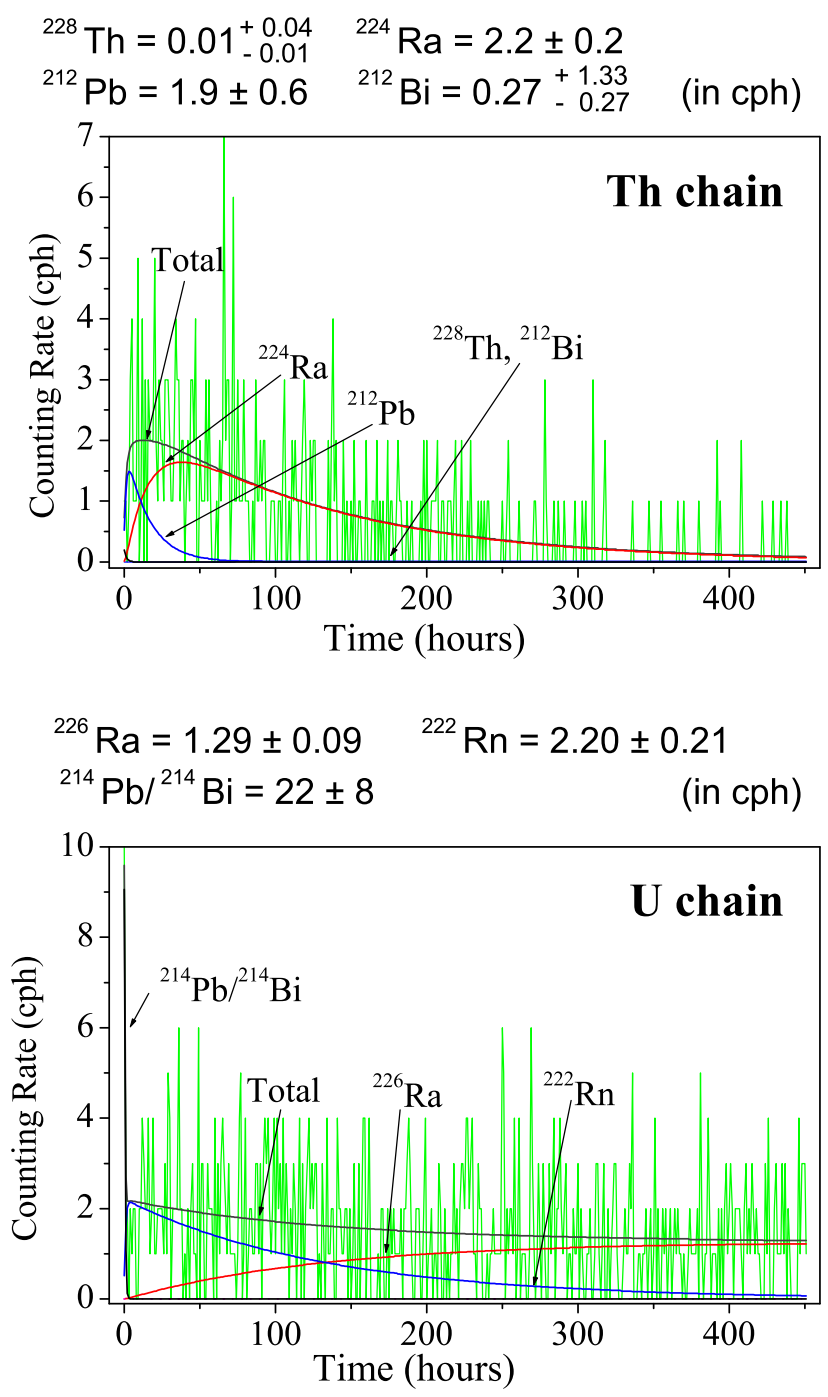

Fig. 4. The fitting of true $\beta-\alpha$ coincidences binned per hour as the counting progressed in time for assay RDA5 for Th and U chain activities. The one sigma errors on the fitted parameters are statistical.

Taking into account the decay of $^{224}$Ra during the transport and processing time of the filters (for RDA5 a factor of 0.78) and all of the relevant efficiencies (Table 1), the fit results for $^{224}$Ra and $^{226}$Ra can be converted to give the activity of each isotope on the filters at the end of extraction (given below in Table 3).

### 4.2. *Disequilibrium within the Thorium and Uranium chains*

Thorium tends to plate out on some surfaces but can form complexes in water, while radium can be leached out of surfaces in contact with the water. As a result it is possible that in an assay $^{228}$Th ($t_{1/2} = 1.9y$) and its daughter $^{224}$Ra ($t_{1/2} = 3.7d$) are not in equilibrium by the time the water has reached the assay column. During the assay any thorium in the water will be extracted with ∼95% extraction efficiency [10], and in the subsequent elution of the radium off the filters very little thorium is removed. So if the filters are left after being eluted after an assay, $^{224}$Ra will grow in from the decay of any extracted $^{228}$Th, and can then be eluted again and the activity measured.

The result for the RDSA9 salt assay, and for an earlier assay (but with less precision) of pure heavy water in a storage tank, was that the $^{228}$Th and $^{224}$Ra activities appeared to be in equilibrium, so that the $^{224}$Ra would be supported during an assay. For the LWA5 and LWA7 light water assays it was found that there was little (<15%) $^{228}$Th activity in comparison with the $^{224}$Ra activity extracted, suggesting that leaching of radium (or plating of thorium, or both) is more significant in the light water where the PMTs are immersed. For the results shown in Table 3, it is assumed that radium is supported in the heavy water assays but not in the light water assays.

Allowing for any decay of $^{224}$Ra during an assay gives the amount in the water provided that there is no significant plating of radium between the cavity and the assay column. Any plating is not expected to be significant with the water flow present in the polypropylene piping and has been measured to be less than 2%. Any contribution of $^{224}$Ra from the piping between the acrylic vessel and the assay filters has been estimated to be less than the equivalent of $0.6\times10^{-15}$g/g of $^{232}$Th [12]. The final results are interpreted as the equivalent amounts of $^{232}$Th and $^{238}$U in heavy water or light water assuming that all the radionuclides are in equilibrium.

As the deuteron photodisintegration rate, which mimics the neutrino NC interaction, depends on the lead isotope activities (see Fig. 1), it is important to avoid any radon contamination both in the heavy water and in the light water. The $N_2$ cover gas together

Table 3
The final results for some of the $D_2O$ and $H_2O$ assays from October 1999 to November 2001; the one sigma errors given are statistical, the systematic errors are discussed in the text.

| Experiment | Date | Tonne | T/m$^2$ | *l*/min/m$^2$ | g Ti/m$^2$/# | $10^{-15}$g Th/g | $10^{-14}$g U/g |
|---|---|---|---|---|---|---|---|
| $D_2O$ target | | | | | | $< 3.8$ | $< 3.0$ |
| RDA2 | 13 Oct 99 | 40 | 8 | 12 | 0.5/H | $< 12$ | $0.28(1\pm0.76)$ |
| RDA3 | 03 Nov 99 | 72 | 15 | 8 | 0.5/H | $7.4(1^{+1.07}_{-1.00})$ | $0.23(1\pm0.37)$ |
| RDA4 | 07 Dec 99 | 209 | 44 | 16 | 0.5/H | $5.0(1\pm0.14)$ | $0.03(1\pm0.42)$ |
| RDA5 | 22 Feb 00 | 361 | 75 | 17 | 0.5/H | $3.6(1\pm0.11)$ | $0.03(1\pm0.19)$ |
| RDSA8 | 30 May 01 | 414 | 100 | 15 | 2.2/P | $2.0(1\pm0.11)$ | $0.28(1\pm0.04)$ |
| RDSA9 | 24 Jul 01 | 185 | 44 | 10 | 2.5/P | $2.2(1\pm0.16)$ | $0.17(1\pm0.05)$ |
| RDSA10 | 25 Sep 01 | 205 | 49 | 19 | 2.5/P | $2.8(1\pm0.14)$ | $0.06(1\pm0.15)$ |
| RDSA11 | 23 Oct 01 | 268 | 64 | 19 | 2.5/P | $1.1(1\pm0.20)$ | $0.03(1\pm0.43)$ |
| $H_2O$ target | | | | | | $< 37$ | $< 45$ |
| LWA4 | 16 May 01 | 11.0 | 21 | 19 | 2.5/P | $21(1\pm0.62)$ | $1.05(1\pm0.66)$ |
| LWA5 | 13 Jun 01 | 18.6 | 36 | 19 | 2.5/P | $79(1\pm0.09)$ | $0.48(1\pm0.25)$ |
| LWA6 | 26 Jun 01 | 34.3 | 66 | 19 | 2.5/P | $58(1\pm0.08)$ | $0.50(1\pm0.30)$ |
| LWA7 | 08 Sep 01 | 27.5 | 53 | 19 | 2.5/P | $85(1\pm0.06)$ | $0.31(1\pm0.17)$ |
| LWA8 | 06 Nov 01 | 42.6 | 82 | 19 | 2.5/P | $57(1\pm0.07)$ | $0.26(1\pm0.26)$ |

# H: Hollow fibre, P: Pleated membrane.

with an adverse temperature gradient near the surface of both the $D_2O$ and $H_2O$ stops any significant radon contamination from mine air. The largest sources of radon in the $D_2O$ are from when the heavy water circulation system is turned on (bringing in radon from assay columns and from permeation through seals) and from when sources are deployed.

## 5. Results and Conclusions

The final results for some representative assays, after subtraction of the measured background for each assay, together with the target limits for $^{232}$Th and $^{238}$U in the light and heavy water for SNO, are given in Table 3; also shown are the amount of water (tonne/m$^2$) and the flow (*l*/min/m$^2$). It is assumed that there was no mixing of returned with sampled water in any of the assays. The activity is assumed to be uniformly distributed throughout the light and heavy water. For each of the assays with the pleated membrane filters, the measured backgrounds were similar and equivalent to a level of $8\times10^{-16}$g/g of $^{232}$Th and of $5\times10^{-16}$g/g of $^{238}$U for an assay of 200 tonnes of heavy water

The one sigma errors given in Table 3 are statistical. The systematic error arises principally through the uncertainty of $\sim$ 24% in the counting efficiency and the efficiencies of the chemical procedures (see Table 1), and to a lesser extent from the uncertainties in the volume of water sampled, in the amount of radium plating and in the background subtraction. The total systematic one sigma errors are estimated to be for the Th chain in $D_2O$ $+26-26$% and in $H_2O$ $+24-24$%; for the U chain in $D_2O$ $+31-31$% and in $H_2O$ $+30-30$%. For the Th chain there is an additional systematic error of $0.6\times10^{-15}$g/g from the limit on any $^{224}$Ra contribution from the piping.

The light water assay LWA8 (see Table 2) shows that good extraction is achieved using the pleated membrane filters loaded with 2.5 g Ti/m$^2$ for $\sim$80 tonnes/m$^2$ and $\sim$20 *l*/min/m$^2$. The assays LWA4-8 suggest that the level of $^{232}$Th is about $50\times10^{-15}$g per gram of light water.

Assuming equilibrium in the decay chain and no plating of radium, it can be seen that the HTiO system has shown that the heavy water has met its limit of $< 3.0\times10^{-14}$g $^{238}$U per gram of heavy water and its limit of $< 3.8\times10^{-15}$g $^{232}$Th per gram of heavy water i.e. the Ra impurities in the heavy water are sufficiently low that the photodisintegration background to the predicted NC signal in the SNO experiment from the standard solar model flux of neutrinos is less than 10%. However, while the light water has easily met its limit for the reduction of radioactivity backgrounds of $< 45 \times 10^{-14}$g $^{238}$U per gram of $H_2O$ it appears to be a little above its limit of $< 37 \times 10^{-15}$g $^{232}$Th per gram of $H_2O$.

The total procedural efficiencies are 24±5% for the Th chain and 32±8% for the U chain. At present the total procedural backgrounds for both Th and U chains are low and at the level of $0.50 \pm 0.07$ and $1.34 \pm 0.26$ counts per hour, respectively. The background errors include a contribution from the systematic error arising from the variation in the background rate from assay to assay. Defining the detection limit as 3 times the standard deviation of the background [19], the HTiO system currently has the sensitivity to measure down to $\sim 3 \times 10^{-16}$g Th/g and to $\sim 3 \times 10^{-16}$g U/g.

## Acknowledgements

The authors would like to thank Professors Werner Rank, Francois Caron and Nelson Belzile for supplying laboratory space in the Chemistry department at Laurentian University and Mick Williams, Tony Handford and other members of the Oxford Mechanical Workshop for making the extraction and elution rigs. We are very grateful to the SNO site UG operations crew and to INCO, Ltd. and their staff at Creighton mine, without whose help this work could not have been conducted, and would like to thank Atomic Energy of Canada, Ltd. (AECL) for the generous loan of the heavy water in cooperation with Ontario Power Generation.

This work was supported in the United Kingdom by the Science and Engineering Research Council and the Particle Physics and Astronomy Research Council; in Canada by the Natural Sciences and Engineering Research Council, the National Research Council, Industry Canada, the Northern Ontario Heritage Fund Corporation, and the Province of Ontario; in the USA by the Department of Energy. Further support was provided by INCO, AECL, Agra-Monenco, Canatom, the Canadian Microelectronics Corporation, AT&T Microelectronics, Northern Telecom, and British Nuclear Fuels, Ltd.